\documentclass[review]{elsarticle}

\usepackage{amsmath}
\usepackage{siunitx, subfig}

\journal{Journal of \LaTeX\ Templates}

\begin{document}

\begin{frontmatter}

\title{Comparing the performance of LDA and GGA functionals in predicting the lattice thermal conductivity of semiconductor materials: the case of AlAs}

\author{Marco Arrigoni }
\ead{marco.arrigoni@tuwien.ac.at}
\author{Georg K. H. Madsen}

\address{Institute of Materials Chemistry, TU Wien, A-1060 Vienna, Austria}

\begin{abstract}
In this contribution we assess the performance of two different exchange-correlation functionals in the first-principle prediction of the lattice thermal conductivity of bulk semiconductors, namely the local density approximation (LDA) and the Perdew-Burke-Ernzerhof implementation of the generalized gradient approximation (GGA). Both functionals are shown to give results in good agreement with experimental measurements. Such a consistency between the two functionals may seem a bit surprising, as the LDA is known to overbind and the GGA to soften the interatomic bonds. Such features ought to greatly affect the value of the system interatomic force constants (IFCs) which are necessary for the first-principle prediction of the lattice thermal conductivity. 
In this study we show that the errors introduced by such approximations tend to cancel themselves. In the case of LDA, the overbinding generates larger absolute third-order IFCs, which tend to increase the three-phonon scattering rates. On the other hand, larger absolute second-order IFCs lead to a a larger acoustic-optical phonon band gap which in turns decrease the available phase space for three-phonon scattering, compensating the increase in the scattering rates due to stiffer IFCs. 
\end{abstract}

\end{frontmatter}

\section{Introduction}
The study of thermal transport in semiconductors has been receiving an increasing amount of consideration  since the last several years. This is not surprising considering the fundamental technological role of semiconductor materials in  power electronic devices and the detrimental effect of high temperatures on the  life time of systems such as high-efficency  LEDs \cite{whiteled}. Moreover, the continuous miniaturization of the device size and the consequent increase  in power loss make heat management an essential task in ensuring the reliability and performance of semiconductor-based transistors \cite{schelling2005}. 

The need to understand the microscopic features of thermal transport has encouraged the development of theoretical approaches, not only with good predictive power, but also able to shed light on the atomistic transport mechanisms and helping in the design and discovery of materials with the desirable thermal properties. 

The first-principle determination of the lattice thermal conductivity is particularly advantageous, given the high predictive power of \emph{ab initio} methods  and the lack of any adjustable parameters, which makes such approaches of general validity and application.   In insulators and semiconductors, heat is transported by lattice vibrations \cite{Peierls1929, Ziman1960}. The first-principle investigation of transport properties in these systems is thus usually based on the \emph{ab initio} determination of the material interatomic force constants (IFCs) and the numerical solution of the phonon Boltzmann transport equation (BTE). Nowadays,  first-principle calculations of bulk semiconductors are routinely employed and different software packages are freely available \cite{shenBTE2014, Togo2015, almaBTE2017}. In addition, the predictive power of such methods is now applicable, not only to ideal bulk crystals, but also to crystals with natural isotope abundances \cite{Ward2009, Lindsay2012}, point defects \cite{Katre2016,Katre2017, Dongre2018,Stern_PRB18,Katre_PRM18,Polanco_PRB18}, alloys \cite{Arrigoni18}, dislocations \cite{Wang2017} and lower dimensional systems such as superlattices \cite{Carrete2018}, nanowires \cite{Carrete2011} and a growing number of other complex systems.

The calculation of the IFCs by first principles is commonly carried out either through the linear-response approach within density-functional perturbation theory \cite{DeCicco1969, Gonze1997, Baroni2001}, or through the finite-displacement method \cite{Kresse1995,Parlinski1997}. 
The finite-displacement method calculates the  IFCs in direct space from the forces induced on the system atoms by displacing some ions from their equilibrium positions. This method requires a supercell expansion of the primitive cell and, if properly used, can reach a similar level of accuracy as the linear-response approach. Moreover it is straightforward and easier to implement and is generally applicable to any system. This method is the approach we employ in this study  in order to calculate the second- and third-order IFCs necessary to determine  the lattice thermal conductivity, $\kappa_\ell$,  from first principles.

As the choice of exchange-correlation functional for calculating the IFCs is concerned, the local density approximation and the Perdew-Burke-Ernzerhof implementations of the generalized gradient approximation (PBE) \cite{Perdew1996}  remain the most common alternatives employed in bulk semiconductors and have shown to be able to accurately predict  $\kappa_\ell$ of these materials \cite{Ward2009, Esfar2011, Lindsay2013, shenBTE2014}. Such accuracy might  appear quite surprising as both LDA and GGA are known for not always being able to correctly reproduce the lattice parameters of solids \cite{Haas2009, He2014}. Experimental measurements of $\kappa_\ell$ at high pressure have shown that this quantity is quite sensitive to the value of the cell parameters \cite{Gerlich1982, Ross1984}. Consequentlye one has to expect that the calculated $\kappa_\ell$ is rather dependent on the chosen exchange-correlation functional. While LDA is known to  overestimate cohesive energies and underestimate cell parameters, the  GGA functionals can sometimes yield an improved description of the system under study, but often tend to overcompensate for the LDA overbinding and overestimate the cell parameters. From the point of view of the material IFCs, the influence of these functionals on the calculated values is therefore opposite in behavior. LDA  predicts stiffer bond from which higher values of the IFCs stems, whereas GGA usually gives softer bonds and thus lower values of the IFCs.

In this contribution, we take $\mathrm{AlAs}$ as a model material for general III-V semiconductor with the zincblende structure and examine the effect of the the exchange-correlation functional on the prediction of $\kappa_\ell$. In particular, we investigate the reasons  behind the  LDA and PBE successful prediction of this quantity for bulk semiconductors.

\section{\emph{Ab Initio} Bulk Lattice Thermal Conductivity}
In the first-principle theory of phonon transport, one usually looks either for the solution of the BTE in the relaxation-time-approximation (RTA) \cite{Ziman1960} or for the iterative solution of the linearized BTE \cite{Omini1995, Omini1996}. 
In the latter, the lattice thermal conductivity tensor is expressed as:
\begin{equation}
\label{eq:BTE}
\boldsymbol{\kappa}_{\ell} = \sum_\lambda c_\lambda \mathbf{v}_\lambda \otimes \mathbf{F}_\lambda,
\end{equation}
where  $\lambda$ is a condensed index for representing a phonon with wave vector $\mathbf{q}$ in the branch $s$: $\lambda \equiv (\mathbf{q}, s)$, $c_\lambda$ is the  contribution of phonon $\lambda$ to the constant volume heat capacity per unit volume,  $\mathbf{v}_\lambda$ is the phonon group velocity and $\mathbf{F}_\lambda \equiv \tau_\lambda (\mathbf{v}_\lambda + \mathbf{\Delta}_\lambda) $ is the converged value of the coefficient of the linear term  in the series expansion of the phonon non-equilibrium distribution, $n_\lambda$, around the equilibrium one, with respect to the  temperature gradient \cite{Omini1995, Omini1996, shenBTE2014}.

 In the expression of $\mathbf{F}_\lambda$, $\tau_\lambda$ represents the phonon lifetime in the RTA. It is thus clear, that setting $\mathbf{\Delta}_\lambda = 0$, reduces the problem of solving the linearized BTE to solving the BTE in the RTA.  The RTA  usually underestimates the value of $\kappa_\ell$ for materials with high lattice thermal conductivity, where the umklapp phonon scattering is weak \cite{Ward2009}. However, it generally agrees with the actual solution of the linearized BTE in semiconductor with lower $\kappa_\ell$ and in particular for III-V semiconductors \cite{Lindsay2013}. For this reason, in this study we will mainly refer to the RTA expression of the BTE, whose form has a more intuitive physical interpretation. In the RTA, equation \eqref{eq:BTE} becomes:
\begin{equation}
\label{eq:BTE-RTA}
\boldsymbol{\kappa}_{\ell} = \sum_\lambda c_\lambda \tau_\lambda \mathbf{v}_\lambda \otimes \mathbf{v}_\lambda,
\end{equation}
In particular, $c_\lambda$ and $\mathbf{v}_\lambda$ are determined by the harmonic properties of the system, as:
\begin{equation}
\label{eq:cv}
c_\lambda = \frac{k_B}{VN} \left( \frac{\hbar \omega_\lambda}{k_B T} \right)^2 n^0_\lambda (n^0_\lambda + 1),
\end{equation} 
\begin{equation}
\label{eq:vg}
\mathbf{v}_\lambda = \nabla_{\mathbf{q}}\omega_\lambda,
\end{equation}
where $k_B$ is the Boltzmann constant, $V$ is the unitcell volume, $N$ is the number of unit cells in the crystal and $T$ is the temperature. $n^0_\lambda$ is the equilibrium distribution of phonon $\lambda$ and $\omega_\lambda$ its frequency, which can be obtained from the second-order IFCs after diagonalizing of the dynamical matrix at the reciprocal space point $\mathbf{q}$.

The scattering rates $1/\tau_\lambda$ arise from the scattering mechanisms in which the phonons of the system are subjected to. For a bulk semiconductor, at temperatures in the order of room temperature, the main scattering processes are those resulting from three-phonon interactions, which can be described by the material third-order IFCs. In particular, the three-phonon scattering rates can be expressed as \cite{Mingo_chapt}:
\begin{equation}
\label{eq:t_3ph}
\frac{1}{\tau^{3ph}_\lambda} = \sum_{\lambda' \lambda''}^+ \Gamma^+_{\lambda \lambda' \lambda''} + \sum_{\lambda' \lambda''}^- \frac{1}{2} \Gamma^-_{\lambda \lambda' \lambda''},
\end{equation}
where $ \Gamma^+_{\lambda \lambda' \lambda''}$ takes into account the processes in which phonon $\lambda$ combines with another phonon giving a third: $\lambda'' \equiv \lambda + \lambda'$ and $ \Gamma^-_{\lambda \lambda' \lambda''}$ the processes in which  phonon $\lambda$ decays in other two phonons: $\lambda'' \equiv \lambda - \lambda'$. The conservation of quasi-momentum requires for the wave vector of phonon $\lambda''$ that $\mathbf{q}'' = \mathbf{q} \pm \mathbf{q}' + \mathbf{Q}$, where $\mathbf{Q}$ is a vector of the crystal reciprocal lattice. The scattering amplitudes, $\Gamma^\pm$, are given by:
\begin{equation}
\label{eq:G+}
\Gamma^+_{\lambda \lambda' \lambda''} = \frac{\hbar \pi}{4} \frac{n^0_{\lambda'} - n^0_{\lambda''}}{\omega_\lambda \omega_{\lambda'} \omega_{\lambda''}} |V^+_{\lambda \lambda' \lambda''}| ^2 \delta(\omega_\lambda + \omega_{\lambda'} - \omega_{\lambda''}),
\end{equation}
\begin{equation}
\label{eq:G-}
\Gamma^-_{\lambda \lambda' \lambda''} = \frac{\hbar \pi}{4} \frac{n^0_{\lambda'} + n^0_{\lambda''} + 1}{\omega_\lambda \omega_{\lambda'} \omega_{\lambda''}} |V^-_{\lambda \lambda' \lambda''}| ^2 \delta(\omega_\lambda - \omega_{\lambda'} - \omega_{\lambda''}),
\end{equation}
where the scattering matrix elements, $V^\pm_{\lambda \lambda' \lambda''}$ are:

\begin{equation}
\label{eq:V+-}
V^\pm_{\lambda \lambda' \lambda''} = \sum_{i \in uc} \sum_{jk}\sum_{\alpha \beta \gamma} \Phi_{\alpha \beta \gamma}(i,j,k) \frac{e_{\alpha, \lambda}(i) e_{\beta, \pm \lambda'}(j)e_{\gamma, -\lambda''}(k)}{\sqrt{M_iM_jM_k}},
\end{equation}
The first sum in equation \eqref{eq:V+-} runs over the atoms in a reference unit cell and the other sums run over all the other atoms in the crystal, $\alpha, \beta \; \mathrm{and} \; \gamma$ run over the Cartesian coordinates,  $\Phi_{\alpha \beta \gamma}(i,j,k)$ is an element of the third-order IFCs tensor, $\mathbf{e}_\lambda(i)$ is the  eigenvector of phonon $\lambda$ projected over the degrees of freedom corresponding to atom $i$ and $M_i$ is the atomic mass of atom $i$.

In addition to the inelastic scattering between phonons, one should also consider the scattering rates arising from the elastic interaction between phonons and the isotopic mass disorder present naturally in any crystal. Such elastic rates, $1/\tau^{iso}_\lambda$, can be calculated using an approximate quantum mechanical formula based on second-order perturbation theory \cite{Tamura1983}. 
The net scattering rates of a bulk semiconductor material are then obtained from the sum of these two contributions:
\begin{equation}
\label{eq:tau}
\frac{1}{\tau_\lambda} = \frac{1}{\tau^{3ph}_\lambda} + \frac{1}{\tau^{iso}_\lambda}.
\end{equation}

We  will not consider in this study the influence of the exchange-correlation functional on the elastic scattering rates, $1/\tau^{iso}_\lambda$, as it was shown by Lindsay \emph{et al.} that their contribution to $\kappa_\ell$  is anyways negligible in $\mathrm{AlAs}$, given the minimal isotope mixtures for $\mathrm{Al}$ and $\mathrm{As}$ \cite{Lindsay2013}. 

\section{Computational Details}
We calculated the lattice thermal conductivity of $\mathrm{AlAs}$ both from the iterative solution of the linearized BTE and from the RTA, using the exchange-correlation functionals LDA and PBE. If not stated otherwise, a sampling mesh of $32\times32\times32$ phonon wave vectors was used to calculate $\kappa_\ell$. The calculations were carried out using the \texttt{almaBTE} computer code \cite{almaBTE2017}.

 For each functional, the primitive cell was completely relaxed until the residual forces on the system were less than \SI{1e-5}{\electronvolt \per \angstrom}. 
The second- and third-order IFCs were obtained from the Helmann-Feynman forces using the finite-displacement method. To minimize the spurious contributions of periodic boundary conditions to the system IFCs, we employed a $5\times5\times5$ supercell expansion of the primitive zincblende cell (250 atoms). The set of atomic displacements necessary for extracting the second-order and third-order force constants were obtained using the software \texttt{Phonopy} \cite{Phonopy} and the \texttt{thirdorder.py} tool of the \texttt{ShenBTE} package \cite{shenBTE2014}, respectively. The magnitude of each displacement was set to \SI{0.02}{\angstrom}. The mixed-space approach of Wang \emph{et al.} \cite{Wang2010} for calculating  the non-analytical term of the dynamical matrix, arising from the the long-range dipole-dipole interactions, was used to obtain correct phonon dispersion relations.

All first principles calculations were done using the code \texttt{VASP} \cite{Kresse1999} within  the projector augmented-wave method \cite{Blochl1994} using a plane-wave cutoff energy of \SI{500}{\electronvolt}. 

\section{Results}
\begin{figure}[bt]
\centering
\includegraphics[width=1\textwidth]{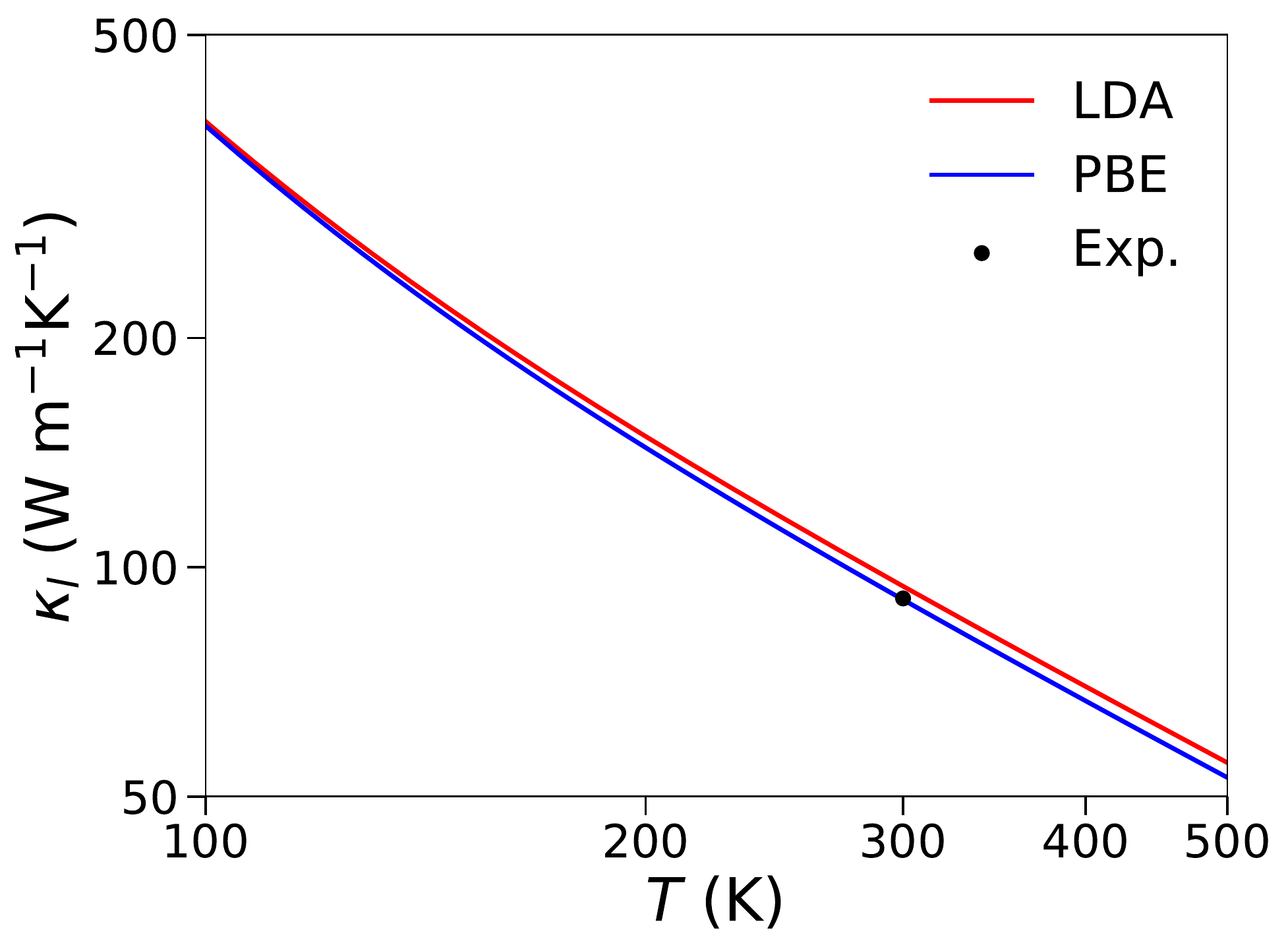} 
\caption{Lattice thermal conductivity obtained from the iterative solution of the linearized BTE for $\mathrm{AlAs}$ using LDA (red) and PBE (blue). The plot is represented in a logarithmic scale.  The experimental data point is taken from Ref. \cite{Afro1973}.}
\label{fig:kl}
\end{figure}

The cell parameter obtained using LDA and PBE has the value \SI{5.64}{\angstrom} and \SI{5.73}{\angstrom}, respectively. As expected, LDA underestimates by $\approx$ 0.4\% the experimental cell parameter of \SI{5.66}{\angstrom} \cite{Lesz1992}, while PBE overestimates it by $\approx$ 1.2 \%. 
 The lattice thermal conductivity of $\mathrm{AlAs}$ calculated with LDA and PBE from the iterative solution of the BTE is plotted in Figure \ref{fig:kl}  as a function of the temperature, where it is also compared with the experimental value at room temperature. The value of $\kappa_\ell$ obtained from the RTA is not shown since it differs by less than 1 \% from the plotted values at around room temperature. 
Taking the room-temperature experimental value \SI{91}{\watt \per \meter \per \kelvin} \cite{Afro1973}, we see that both LDA and PBE yield very accurate results. The former predicts the room temperature $\kappa_\ell$ at $\approx$ \SI{94}{\watt \per \meter \per \kelvin}, while  the latter at  $\approx$ \SI{91}{\watt \per \meter \per \kelvin}. These values differ from each other by approximately the 3 \% and are both in very good agreement with the experimental measurement.

\begin{figure}[bt]
\centering
\includegraphics[width=1\textwidth]{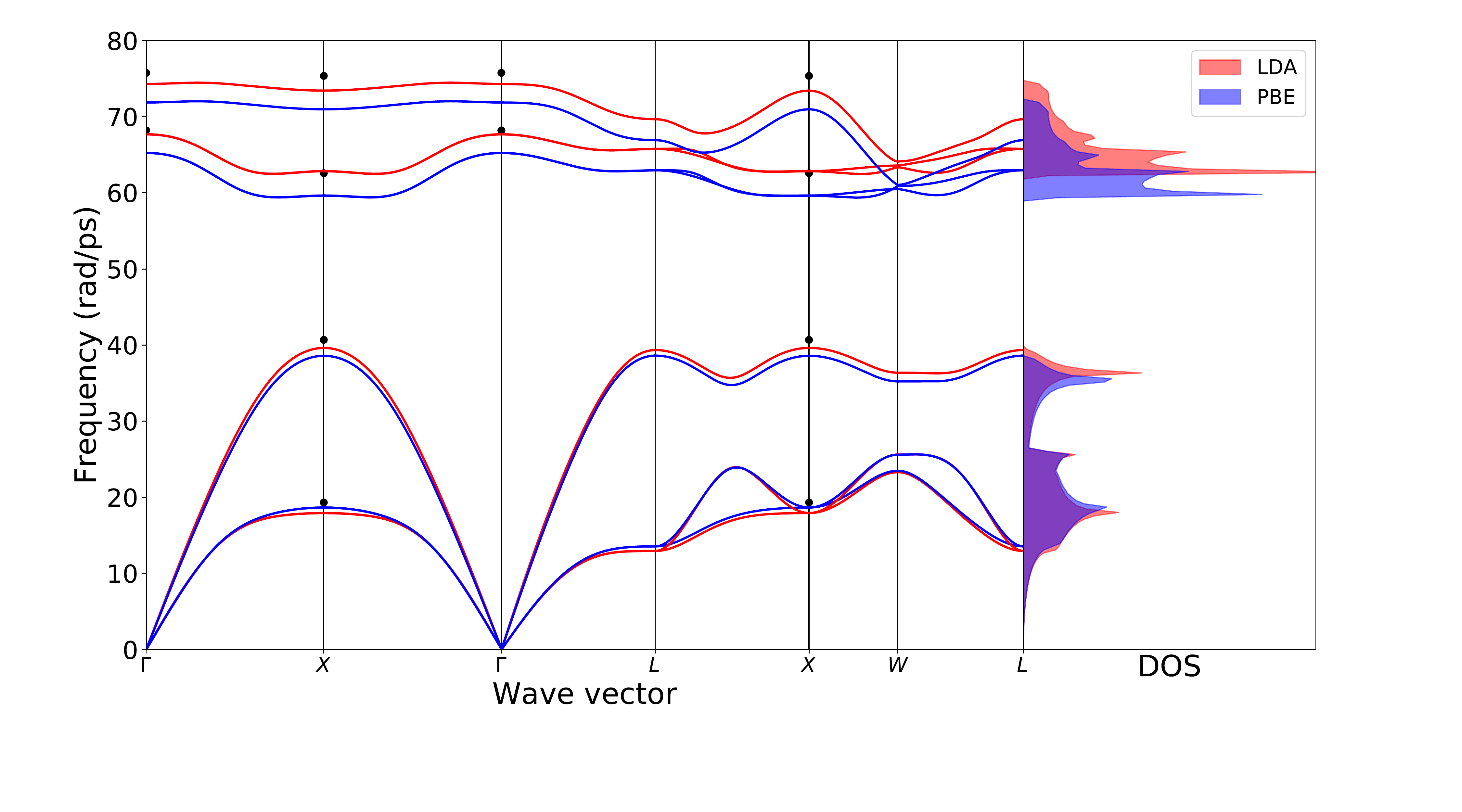} 
\caption{Phonon dispersion relations and density of states calculated within the harmonic approximation using the LDA (red) and PBE (blue) exchange-correlation functionals. The dots represent experimental data points taken from Ref. \cite{LB1}.}
\label{fig:bands}
\end{figure}

To investigate the influence of the exchange-correlation functional, and in particular of the predicted lattice constants, on $\kappa_\ell$, we start with the analysis of the crystal harmonic properties.
The stiffer chemical bonds predicted by LDA lead to higher phonon frequencies, whereas the softer bonds described by PBE are responsible for lower frequencies. These observations are summarized by Figure \ref{fig:bands} which displays and compares the phonon dispersion relations calculated with LDA and PBE second-order IFCs. As the discrepancy between the calculated and experimental cell parameter is larger in PBE, the agreement between the calculated and experimental phonon frequencies is worse than in LDA.
From the picture one can also see that, while acoustic phonon frequencies are only slightly affected by the choice of the functional, optical phonons are rigidly shifted downwards  in PBE as compared with LDA.  As a net effect, the gap between acoustic and optical modes is reduced by approximately \SI{3}{\radian \per \pico \second} in PBE  as compared to LDA.

\begin{figure}[bt]
  \centering
  \subfloat[]{ \includegraphics[width=0.45\textwidth]{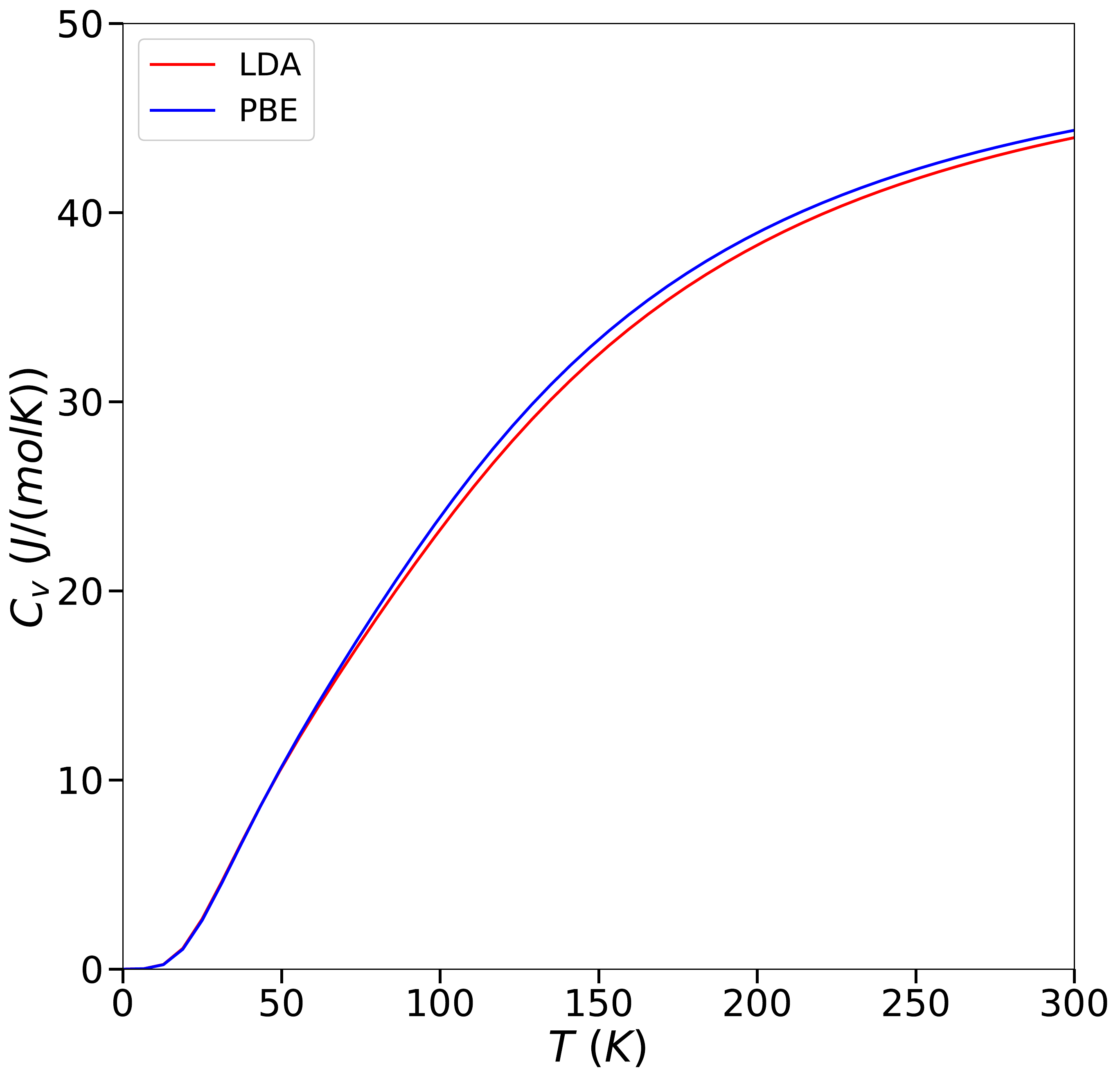}  \label{fig:cv}}  
  \hfill
  \subfloat[]{ \includegraphics[width=0.45\textwidth]{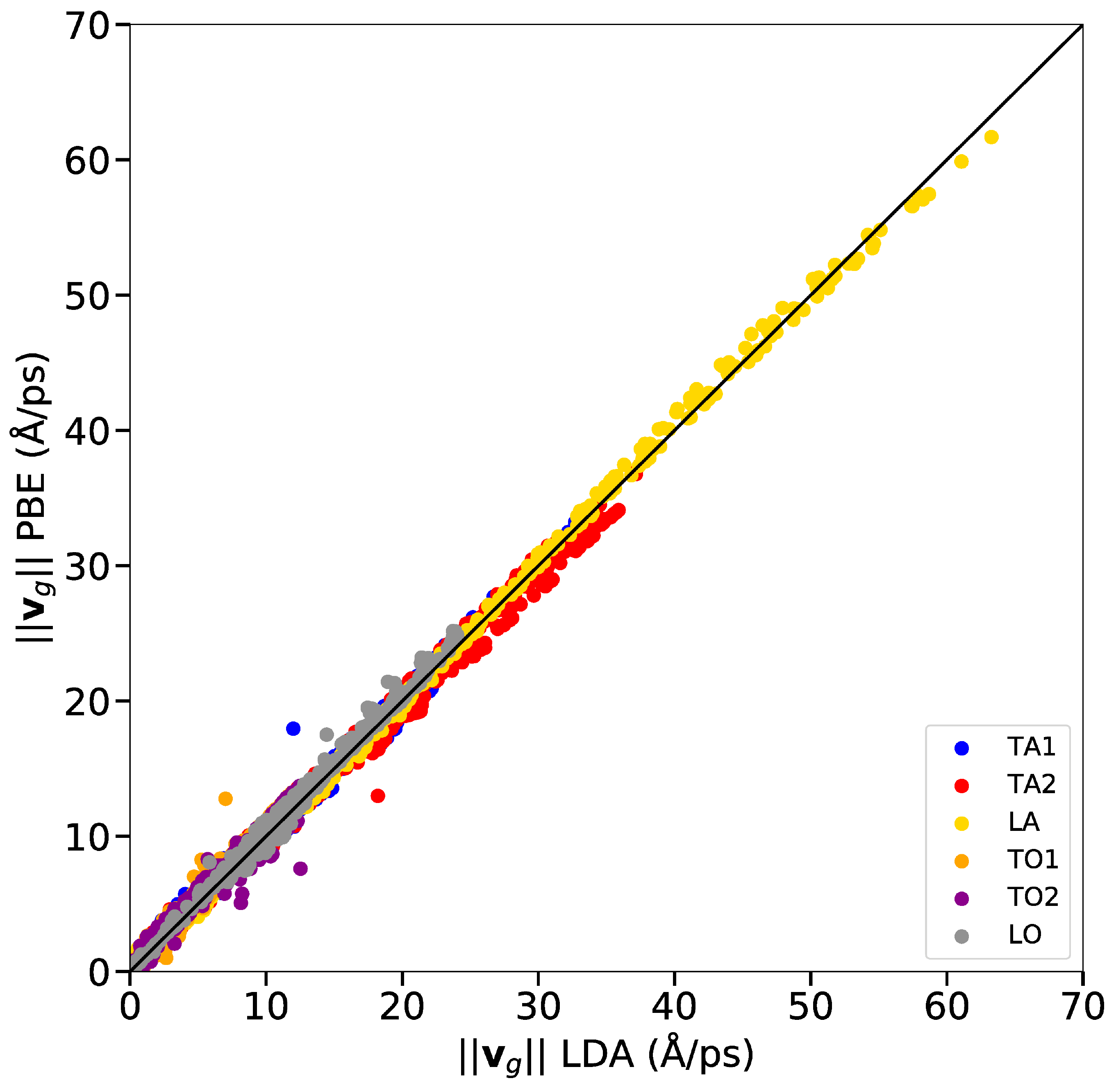} \label{fig:velocities}}

 \caption{(a) Molar constant volume heat capacity of $\mathrm{AlAs}$ calculated with LDA (red) and PBE (blue) as a function of the temperature. (b) Comparison of the phonon group velocities of $\mathrm{AlAs}$ calculated with LDA and PBE. Different colors represent different phonon branches.}
  
\end{figure}

On the other hand, the discrepancy in the LDA and PBE predicted lattice constant has only a very marginal effect on the system constant volume heat capacity, $C_\mathrm{v}$, and phonon group velocities. Figure \ref{fig:cv} and \ref{fig:velocities} shows a comparison between these two quantities calculated  using LDA or PBE. As one can see,  LDA and PBE predict an almost identical constant volume heat capacity. Also the group velocities are rather insensitive to the exchange-correlation functional. PBE predicts somewhat larger group velocities for the longitudinal acoustic modes and optical modes, while LDA predicts slightly larger group velocities for transverse acoustic phonons.  The small influence of the different predicted cell parameters on the group velocities is not surprising since from Figure \ref{fig:bands} it is evident that the effect of the exchange-correlation functional is to rigidly shift the phonon frequencies without affecting noticeably the slope of the dispersion relations. 

\begin{figure}[bt]
\centering
\includegraphics[width=1\textwidth]{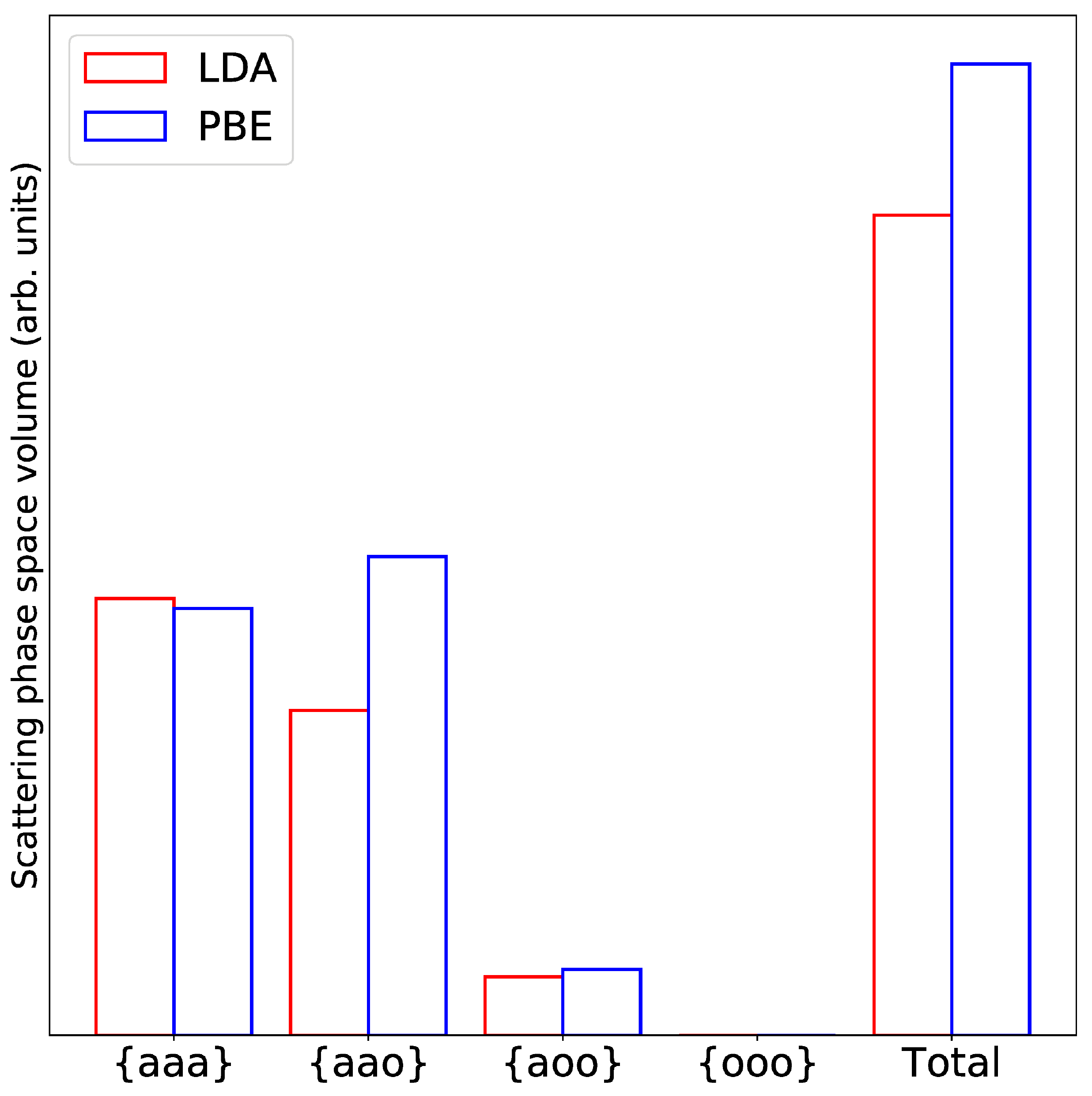} 
\caption{Available phase-space volume for three-phonon scattering processes calculated using LDA and PBE and represented for all possible events involving the various phonon  branches.}
\label{fig:volumes}
\end{figure}

While the rigid shift of the optical modes frequencies has a minor effect on $C_\mathrm{v}$ and the group velocities, it has a more important influence on the three-phonon scattering rates. To understand this we have to consider equations \eqref{eq:G+} and \eqref{eq:G-}. We see that the scattering amplitudes are constrained by the energy conservation conditions: $\hbar \omega_{(\mathbf{q}'', s'')} = \hbar \omega_{(\mathbf{q}, s)} \pm \hbar \omega_{(\mathbf{q}', s')}$ and by the crystal momentum conservation conditions: $\mathbf{q}'' + \mathbf{Q} = \mathbf{q} \pm \mathbf{q}'$. Such constraints severely limit the number three-phonon scattering events and are taken into account in equations \eqref{eq:G+} and \eqref{eq:G-} by the Dirac's deltas $\delta(\omega_{(\mathbf{q}, s)} \pm \omega_{(\mathbf{q}', s')} - \omega_{(\mathbf{q} \pm \mathbf{q}'-\mathbf{Q}, s'')})$.
These define an hypersurface, or volume, in phase space, determined by the equation $\hbar \omega_{(\mathbf{q}, s)} = \hbar \omega_{(\mathbf{q} \pm \mathbf{q}'-\mathbf{Q}, s'')} \mp \hbar \omega_{(\mathbf{q}', s')}$.  This observation suggests to define the phase space available for the three-phonon scattering involving phonon $(\mathbf{q}, s)$ by \cite{Ziman1960, Lindsay2008}:
\begin{equation}
\label{eq:3phV}
D^\pm(\mathbf{q}, s) = \sum_{s', s''}\int_{\mathbf{q}'} \delta(\omega_{(\mathbf{q}, s)} \pm \omega_{(\mathbf{q}', s')} - \omega_{(\mathbf{q} \pm \mathbf{q}'-\mathbf{Q}, s'')}) \, d\mathbf{q}',
\end{equation}
where the integration is over the volume of the first Brillouin zone.
The net phase space volume available for three-phonon scattering is obtained from:
\begin{equation}
P_3 = (P_3^+ + \frac{1}{2} P_3^-),
\end{equation}
where $P_3^\pm = \sum_s \int_\mathbf{q} D^\pm(\mathbf{q}, s) \, d\mathbf{q}$.
The larger $P_3$, the more three-phonon processes are allowed and the larger $1/\tau^{3ph}$ becomes. Indeed it has been shown that, for those group IV, III-V and II-VI semiconductors materials where thermal transport is dictated by three-phonon scattering, the lattice thermal conductivity  varies inversely with the magnitude of $P_3$ \cite{Lindsay2008}.
By fixing the branch indexes $s, s'$ and $s''$ in equation \eqref{eq:3phV}, we can similarly define the three-phonon scattering volume available for phonons in selected branches. The magnitude of such volumes are represented in Figure \ref{fig:volumes}, where the values obtained from PBE and LDA are compared.
 The letters ``$a$'' and ``$o$'' indicate that any of the considered three phonons is an acoustic or optical one, respectively. Thus, \emph{e.g.}, the symbol $\{aao\}$ indicates that the phase space volume available for the scattering between two acoustic and one optical phonon is considered. 
It is clear that the reduced acoustic-optical gap predicted by PBE with respect to LDA is responsible for an increased number of scattering events involving two acoustic phonons and an optical phonon, which noticeably increases the total three-phonon scattering volume. 

\begin{figure}[bt]
\centering
\includegraphics[width=1\textwidth]{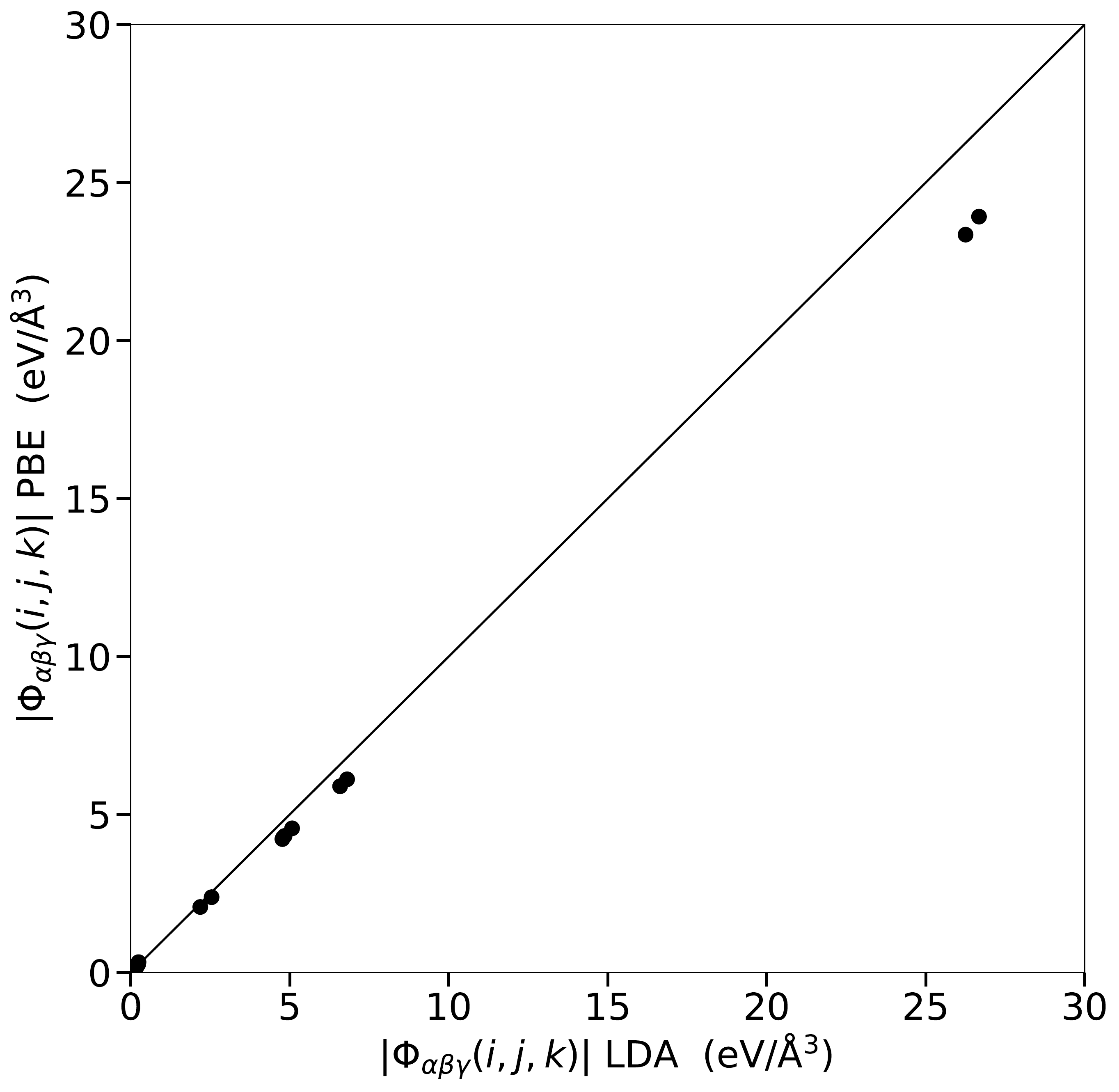} 
\caption{Comparison between the symmetry-irreducible third-order IFCs as predicted by LDA and PBE for $\mathrm{AlAs}$. }
\label{fig:ifc3}
\end{figure}

The three-phonon scattering volume is an important predictor for $1/\tau^{3ph}$ but it is not the only relevant contribution to the scattering rates. As shown in equations \eqref{eq:G+} and \eqref{eq:G-}, the scattering amplitudes are also affected by the value of the scattering matrix elements given by equation \eqref{eq:V+-}, which, in turn, depend on the magnitude of the third-order IFCs. In Figure \ref{fig:ifc3} we compare the magnitude of the third-order IFCs as obtained from PBE and LDA. 
As PBE describes softer bonds than LDA, it is not surprising that the magnitudes of the third-order IFCs are larger in the latter than in the former. This is especially true if we consider the IFCs with largest absolute value. The decrease in the third-order IFCs calculated by PBE, yields lower scattering matrix elements and tends to compensate the increase in three-phonon scattering volume.

\begin{figure}[bt]
\centering
\includegraphics[width=1\textwidth]{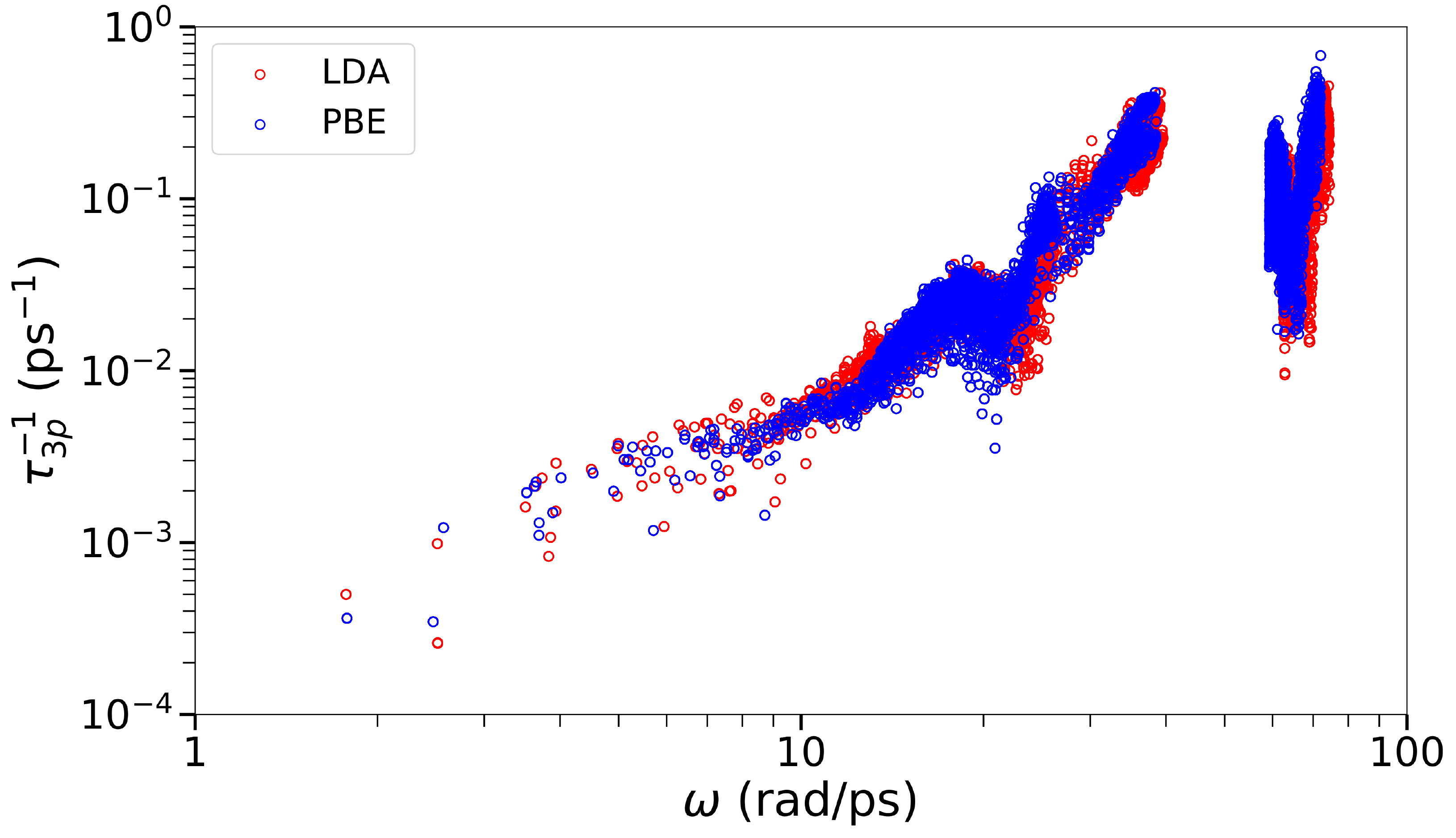} 
\caption{Comparison of the room-temperature three-phonon scattering rates calculated by LDA and PBE for $\mathrm{AlAs}$. }
\label{fig:w3}
\end{figure}

Softer bonds thus induce a smaller acoustic-optical phonon gap, thereby increase the number of accessible three-phonon scattering events, and smaller IFCs, which in turn decrease the magnitude of the matrix scattering elements. The net effect on the scattering rates is shown in Figure~\ref{fig:w3}, which compares the $1/\tau^{3ph}$ calculated with LDA and PBE as a function of the phonon frequency. As one can see the rates assumes values very close to each other, and since $c_\lambda$ and $\mathbf{v}_\lambda$ assume similar values in both LDA and PBE, also the value of $\kappa_\ell$ predicted by these functional is very similar. 

\section{Conclusions}
We have performed first-principles calculations of the IFCs of bulk $\mathrm{AlAs}$ within the LDA and PBE approximations in order to evaluate the effect of the exchange-correlation functional on the lattice thermal conductivity. LDA  underestimates the crystal lattice parameters, yielding stiffer IFCs, whereas PBE overestimates the cell parameters and gives softer IFCs. Notwithstanding these shortcomings, both functionals are able to predict $\kappa_\ell$ in a very good agreement with the experimental data. The reasons behind this behavior can be summarized as follows: the main effect of the softening of the IFCs affecting the PBE calculations is to reduce the gap between the acoustic and optical phonon branches. This, in turn, increase noticeably the number of possible scattering events involving two acoustic and one optical phonons. At the same time, softer third-order IFCs tends to give rise to smaller scattering matrix elements, which  tends to compensate for the increased number of scattering events. As a net effect, the three-phonon scattering rates predicted by PBE and LDA are comparable, leading to values of $\kappa_\ell$ in agreement with each other.

\section{Acknowledgements}
The authors thank Dr. J. Carrete for the useful discussions and comments.
The authors acknowledge support from the European Union's Horizon 2020 Research and Innovation Action, grant number 645776 (ALMA) and from the Austrian Science Fund (FWF), project CODIS (I 3576-N36).  We also thank the Vienna Scientific Cluster for providing the computational facilities (project number 70958: ALMA). 

\section*{Data availability}
The raw and processed data required to reproduce these findings are available to download from:

 https://data.mendeley.com/datasets/7tvtbgjpdm/1

\end{document}